\documentclass[twocolumn,10pt,amsmath,amssymb,aps,pra,superscriptaddress]{revtex4-1}
\usepackage{graphicx}
\usepackage{color}
\usepackage{bm}
\usepackage{amssymb,amsmath,booktabs}
\usepackage[dvipsnames]{xcolor}
\usepackage{enumitem}
\usepackage[caption=false]{subfig}
\usepackage{hyperref}

\begin{document}

\title{Photon-Mediated Charge Exchange Reactions Between $^{39}$K Atoms and  $^{40}$Ca$^+$ Ions in a Hybrid Trap}

\author{Hui Li}
\affiliation{Department of Physics, Temple University, Philadelphia, Pennsylvania 19122, USA}
\author{S. Jyothi}
\affiliation{Department of Electrical and Computer Engineering, Duke University, Durham, North Carolina 27708, USA}
\author{Ming Li}
\affiliation{Department of Physics, Temple University, Philadelphia, Pennsylvania 19122, USA}
\author{Jacek K{\l}os}
\affiliation{Department of Physics, Temple University, Philadelphia, Pennsylvania 19122, USA}
\affiliation{Department of Chemistry and Biochemistry, University of Maryland, College Park, Maryland 20742, USA}
\author{Alexander Petrov}
\affiliation{Department of Physics, Temple University, Philadelphia, Pennsylvania 19122, USA}
\affiliation{NRC, Kurchatov Institute PNPI, Gatchina, 188300, and Division of Quantum Mechanics, St.~Petersburg State University, St.~Petersburg, 199034, Russia}
\author{Kenneth R Brown} 
\affiliation{Department of Electrical and Computer Engineering, Duke University, Durham, North Carolina 27708, USA}
\author{Svetlana Kotochigova}
\email{skotoch@temple.edu}
\affiliation{Department of Physics, Temple University, Philadelphia, Pennsylvania 19122, USA}

\date{\today}

\begin{abstract}
We present experimental evidence of charge exchange between laser-cooled potassium $^{39}$K atoms and calcium $^{40}$Ca$^+$ ions in a hybrid atom-ion trap and give quantitative theoretical explanations for the observations. The $^{39}$K atoms and $^{40}$Ca$^+$ ions are held in a magneto-optical (MOT) and a linear Paul trap, respectively. Fluorescence detection and high resolution time of flight mass spectra for both species are used to determine the remaining number of $^{40}$Ca$^+$ ions, the increasing number of  $^{39}$K$^+$ ions, and $^{39}$K number density as functions of time. 
Simultaneous trap operation is guaranteed by alternating periods of MOT and $^{40}$Ca$^+$ cooling lights, thus avoiding direct ionization of $^{39}$K by the $^{40}$Ca$^+$ cooling light.  
We show that the K-Ca$^+$ charge-exchange rate coefficient increases linearly from zero with $^{39}$K number density and the fraction of $^{40}$Ca$^+$ ions in the 4p\,$^2$P$_{1/2}$ electronically-excited state. Combined with our theoretical analysis, we conclude that these data can only be explained by a process that starts with a potassium atom in its electronic ground state and a calcium ion in its excited 4p\,$^2$P$_{1/2}$ state producing ground-state $^{39}$K$^+$ ions and metastable, neutral Ca\,(3d4p$^3$P$_1$) atoms, releasing only 150 cm$^{-1}$ equivalent relative kinetic energy. Charge-exchange between either ground- or excited-state $^{39}$K and ground-state $^{40}$Ca$^+$ is negligibly small as no energetically-favorable product states are available. Our experimental and theoretical rate coefficients are in agreement given the uncertainty budgets.
\end{abstract}

\maketitle

\section{Introduction}
Over the past few decades, laser cooling and trapping of atoms, ions, and molecules have significantly contributed to advancing fundamental atomic, molecular, and optical physics research. More recently,  experiments successfully  merged  ion  and neutral atom trapping techniques. These experiments include the study of  reaction dynamics
of neutral atoms and atomic  ions \cite{Grier2009,Zipkes2010,Schmid2010,Hall2011,Rellergert2011,Harter2012,Ravi2012,Sivarajah2012,Ray2014,Smith2014,Goodman2015,Meir2016,Saito2017,Joger2017,Sikorsky2018,Kwolek2019,Tomza2019}, as well as of cold atoms and molecular ions\cite{Hall2012,Deiglmayr2012,Sullivan2012,Rellergert2013}. Here, the long storage times of laser-cooled ion-atom systems  has allowed the study of processes that are difficult to observe otherwise.  The internal and external states of the atoms and ions can be manipulated using laser fields, which enables a careful investigation of the quantum-state dependence of a  process.

Most research has focused on colliding neutral and ionic atoms or molecules prepared in their electronic ground state. When one or both colliding partners, however, are prepared in excited electronic states  more inelastic channels become available and richer dynamics occurs with  corresponding challenges in the theoretical interpretation. Recently, such inelastic collisions have been reported for the ion in an excited state \cite{Staanum2004,Hall2011,Ratschbacher2012,Saito2017,Kwolek2019,Ben2019} and for the excited atom colliding with an ion or a charged molecule \cite{Mills2019, Li2019,Kwolek2019,Puri2019,Dorfler2019}, or both particles in an excited state \cite{Kwolek2019}. 
In all these studies the role of excitation in collisional dynamics of various atomic and ionic species has been investigated. However, the control of reaction pathways and population of the excited states is also critically important for 
multiple applications of ions such as optical frequency standards, quantum simulation, quantum computing and astronomy \cite{Schmidt2003, Margolis2004, Kreuter2005, Haffner2008, Blatt2012}.

Here, we report on the theoretical and experimental investigation of  cold charge exchange  collisions between  laser-cooled potassium ($^{39}$K) atoms and calcium ($^{40}$Ca$^+$) ions in the ground and excited states under controlled experimental conditions. The nearly-equal masses of potassium and calcium  enables the simultaneous trapping of the reactant and resultant ions in the same ion trap. Further, direct detection of the ions using a high-resolution time-of-flight mass spectrometer (TOFMS) allows us to  follow charge exchange  as a function of time. We also show that our data are explained by
a process that starts with a potassium atom in its electronic ground state and a calcium ion in its excited 4p\,$^2$P$_{1/2}$ state producing ground-state $^{39}$K$^+$ ions and metastable, neutral Ca\,(4s4p$^3$P$_1$) atoms, releasing only 150 cm$^{-1}$ equivalent relative kinetic energy.
Our measured charge-exchange rate coefficient is in good agreement with the theoretical estimate.

The article is organized as follows. In section II, we describe the experimental setup and measurements of the charge-exchange rate coefficient. The theoretical models for different charge exchange pathways are detailed in section III, followed by a discussion in section IV. Details of our electronic structure calculation is given in Supplemental Material.

\section{Experimental Setup and Observations}

\begin{figure}
\includegraphics[width=\columnwidth,trim=0 0 0 0,clip]{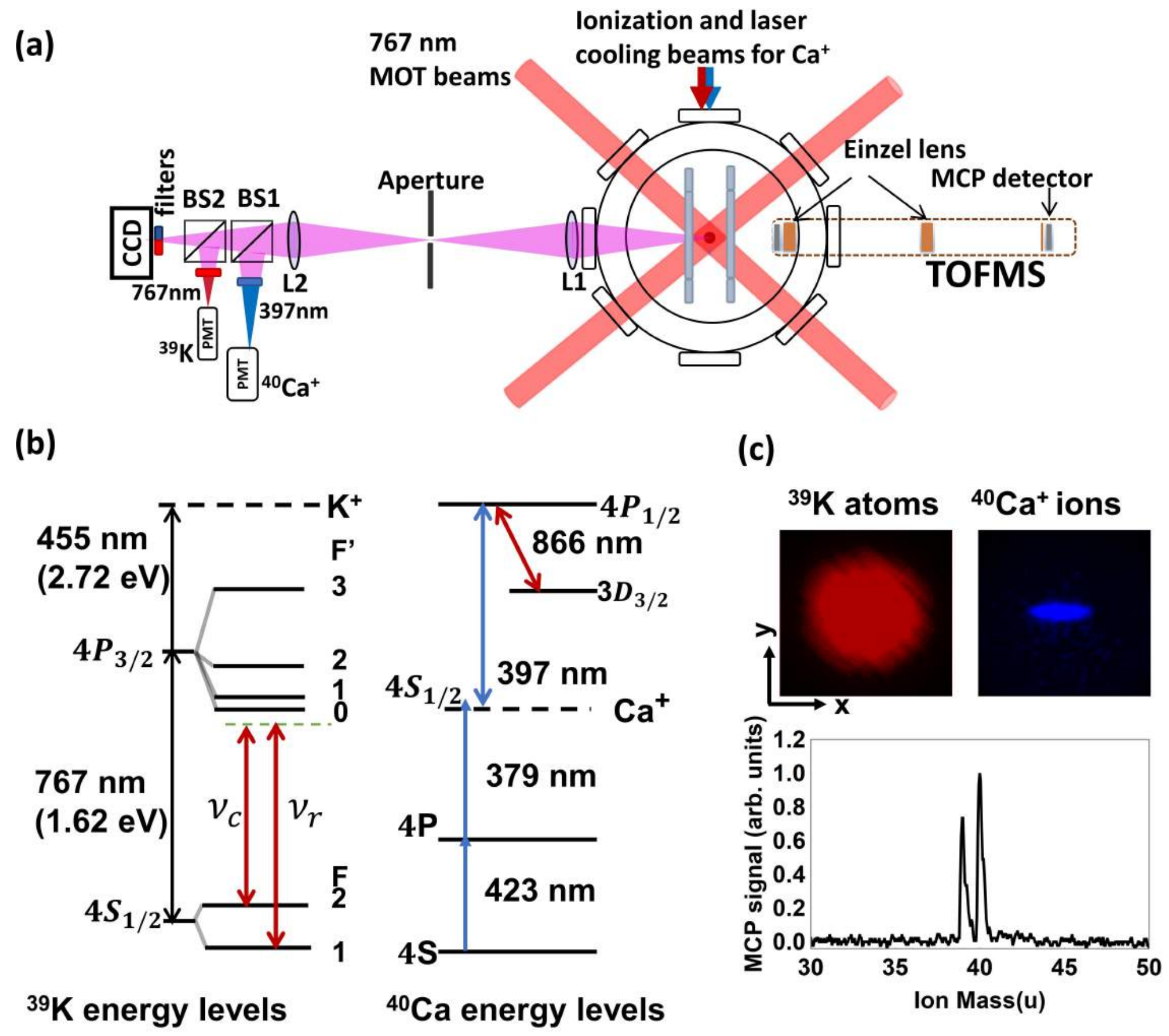}
\caption{\label{fig:expt}(a) Top view of the experimental apparatus illustrating laser setups, fluorescence detection, and ion time of flight detection. (b) the relevant states, energy levels, and laser colors (not to scale) of $^{39}$K, $^{39}$K$^+$, $^{40}$Ca, and $^{40}$Ca$^+$. 
Red vertical lines with arrows and labels $\nu_{\rm c}$ and $\nu_{\rm r}$ correspond to the cooling and repumper lasers
for $^{39}$K, respectively.  The four lasers (blue and red lines with arrows) that control neutral $^{40}$Ca and ionic  $^{40}$Ca$^+$ are discussed in the text.
(c) CCD images of the $^{39}$K MOT and $^{40}$Ca$^+$ ions, and a TOF mass spectrum showing both $^{39}$K$^+$ and $^{40}$Ca$^+$ ion peaks.}
\end{figure}

The experiments are performed in a spatially overlapped ion-atom hybrid trap comprised of a magneto optical trap (MOT) for neutral $^{39}$K atoms and a linear Paul trap for $^{40}$Ca$^+$ ions \cite{Jyothi2019}.
A schematic of the experimental setup is shown in Fig.~\ref{fig:expt}(a). The ion trap is a segmented electrode linear quadrupole trap with four central rf electrodes (driven at 1.7 MHz) for radial trapping and eight end cap dc electrodes for axial trapping. The apparatus is integrated with a high-resolution TOFMS consisting of two Einzel lenses for ion collimation and a microchannel plate for ion detection.The ions are radially extracted from the Paul trap towards the detector by turning off the rf voltages while, simultaneously, applying high dc voltages on the central electrodes to create a potential gradient.

Neutral $^{39}$K atoms derived from a potassium ampule are cooled and trapped in a MOT using three mutually orthogonal and retro-reflected laser beams. The energy-level diagram of $^{39}$K atoms relevant for our experiments is shown in Fig.~\ref{fig:expt}(b).  Cooling and repumper beams are derived from the same 767 nm laser by MOGLabs  \cite{moglabs_laser} locked to the D2 crossover peak of a saturation absorption spectrum. A pair of acousto-optic modulators (AOMs) are used to set the desired detunings for the cooling (20 MHz to the red of the  4s($F=2$) to 4p($F’=3$) transition) and repumper (12 MHz to the red of the $F=1$ to $F’=2$ transition) beams. The quadrupole magnetic field for the MOT is created by a pair of coils in anti-Helmholtz configuration mounted outside the vacuum chamber.

Neutral calcium atoms are produced by heating a calcium dispenser with a current of about 2.7 A. The $^{40}$Ca atoms are ionized by two-photon ionization using  lasers at wavelengths of 423 nm and 379 nm focused at the center of the ion trap. The $^{40}$Ca$^+$ ions are Doppler cooled using a 397 nm laser and repumped with an 866 nm laser. The relevant states and energy levels of $^{40}$Ca and $^{40}$Ca$^+$ are shown in Fig.~\ref{fig:expt}(b). All  laser beams for $^{40}$Ca$^+$ are derived from a multi-diode laser module from Toptica Photonics \cite{toptica_laser}.

To study cold interactions between $^{39}$K  and $^{40}$Ca$^+$, the centers of the MOT and Paul trap must coincide. 
Optimization of the trap overlap is achieved by either moving the MOT center by changing the currents through the coil pair or by moving the center of the Paul trap  by changing the dc voltages on the ion trap electrodes. The quality of the relative positioning of the trap centers has been presented in detail in \cite{Jyothi2019}. For stable co-trapping and cooling of $^{39}$K and $^{40}$Ca$^+$, the 767 nm and 397 nm laser beams are alternately switched at a frequency of 2 kHz using AOMs \cite{Jyothi2019}. This prevents ionization of potassium atoms in the MOT from the 4P$_{3/2}$ state by the 397 nm Ca$^+$ cooling laser beam and  limits the reactants collision channels to K(S)+Ca$^+$(P), K(P)+Ca$^+$(S), and K(S)+Ca$^+$(S).

Fluorescence from $^{39}$K and $^{40}$Ca$^+$ is detected using  photo multiplier tubes (PMT) and an EM-CCD camera. 
The $^{39}$K number density is determined from the fluorescence signal.
Ions are also detected using  TOFMS, enabling the direct observation of non-fluorescing reaction products such as the $^{39}$K$^+$ ions. With the optimal extraction voltages, a mass resolution $m/\Delta m$ of 208 is achieved, which is sufficient to resolve the $^{40}$Ca$^+$ and $^{39}$K$^+$ ion peaks.  EM-CCD images of $^{39}$K  and $^{40}$Ca$^+$  ions as well as a representative TOF mass spectrum of $^{39}$K$^+$ and $^{40}$Ca$^+$ ions are shown in Fig.~\ref{fig:expt}(c).

\begin{figure}
\includegraphics[width=\columnwidth,trim=0 130 0 120,clip]{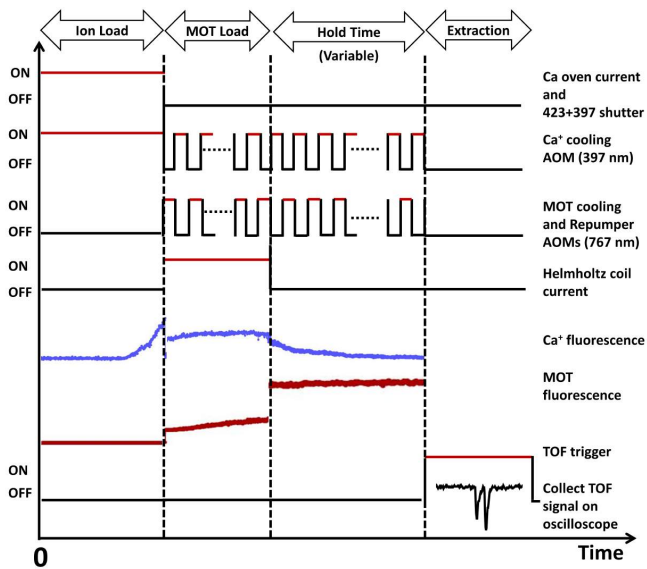}
\caption{Experimental control sequence to measure the charge-exchange rate between $^{39}$K and $^{40}$Ca$^+$ as described in the text. Since the 767 nm and 397 nm laser beams are never simultaneously turned on,  excited $^{39}$K  and excited $^{40}$Ca$^+$  are not simultaneously present. Consequently, charge-exchange collisions only occur  via the K(S)+Ca$^+$(P), K(P)+Ca$^+$(S), and K(S)+Ca$^+$(S) channels. 
}
\label{fig:timing}
\end{figure}

The  experiment is controlled by a Kasli FPGA and programmed using ARTIQ (Advanced  Real-Time  Infrastructure  for  Quantum  physics)  \cite{Bourdeauducq2018, artiq, Jyothi2019} provided by m-labs. A typical experimental sequence to measure  $^{39}$K-$^{40}$Ca$^+$ charge exchange is shown in Fig.~\ref{fig:timing}. First, we load and cool approximately 1000 $^{40}$Ca$^+$  into the ion trap. The photo-ionizing lasers for calcium are then blocked using a mechanical shutter to avoid ionization of $^{39}$K atoms by these lasers. The MOT laser beams are now turned on to load the MOT to saturation. While loading the MOT, currents on an additional pair of magnetic coils in Helmholtz configuration are turned on to shift the MOT magnetic-field center a few mm from the center of the ion trap. This  avoids any interaction between $^{39}$K  and $^{40}$Ca$^+$  during the MOT loading process. The MOT center is then moved back to the ion-trap center by turning off the Helmholtz coil current. The $^{39}$K atoms and $^{40}$Ca$^+$ ions are then allowed to interact for different hold times. 
During the hold time, the 767 nm and 397 nm laser beams are never on at the same time. 

\begin{figure}
\includegraphics[width=\columnwidth,trim=0 0 0 0,clip]{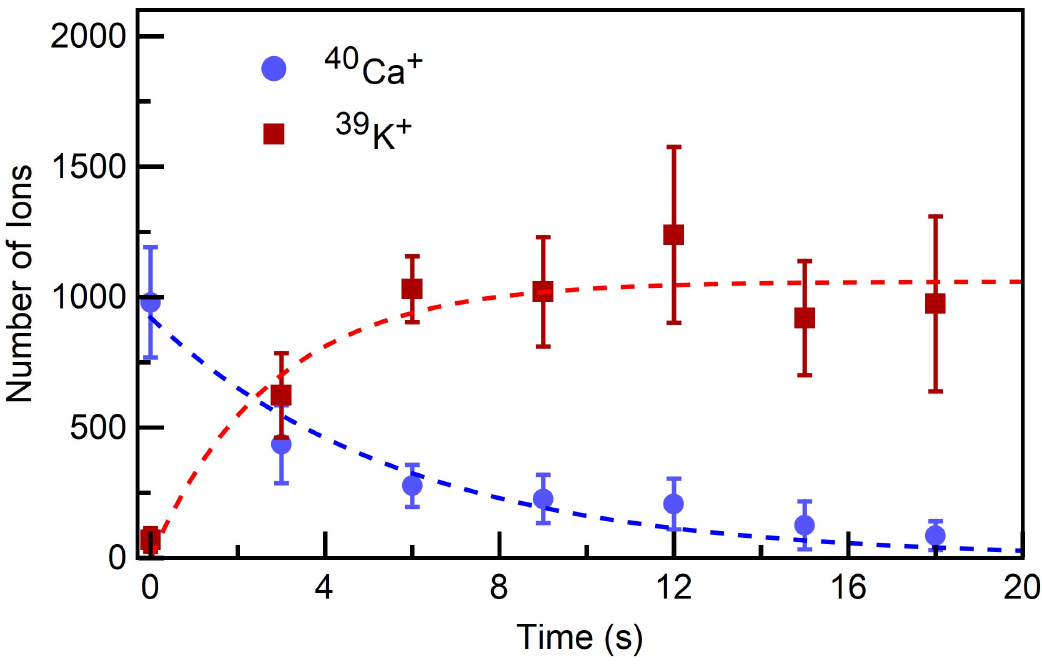}
\caption{\label{fig:TOF}  The number of $^{40}$Ca$^+$  and $^{39}$K$^+$ ions as functions of time for which the $^{40}$Ca$^+$ ions and $^{39}$K atoms are held together. Error bars are one standard deviation uncertainties based on six  measurements. Dashed lines are exponential fits to the data.
The $^{39}$K number density in the MOT is $10^9$ atoms/cm$^3$.}
\end{figure}

 \begin{figure}
\includegraphics[width=\columnwidth,trim=0 0 0 0,clip]{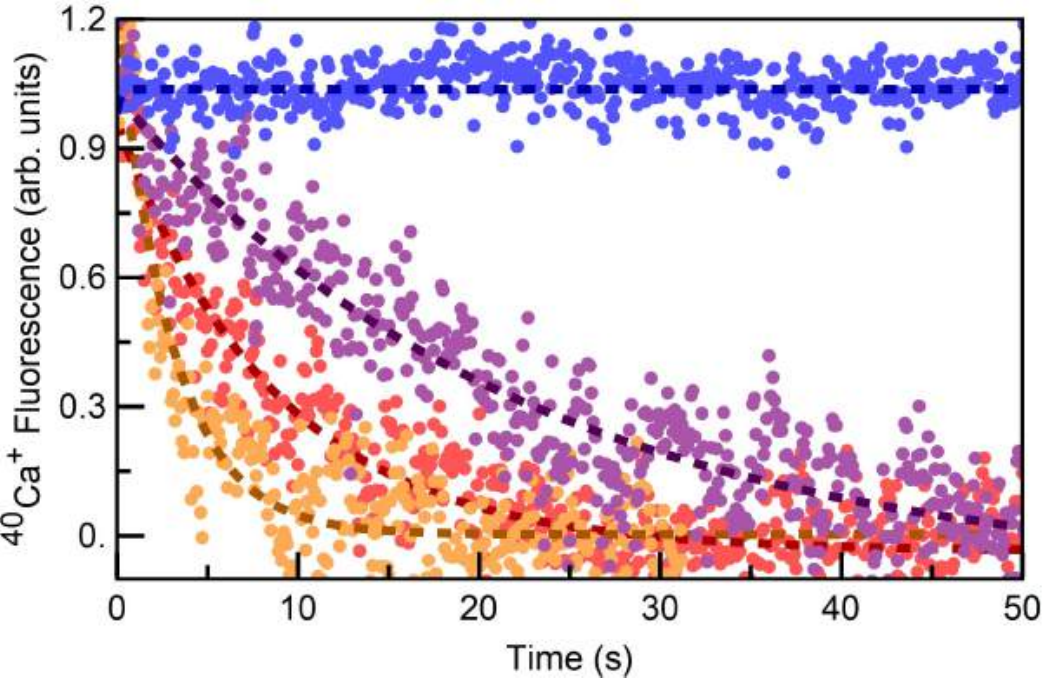}
\caption{\label{fig:flr} Typical fluorescence data from $^{40}$Ca$^+$ ions as functions of time in the absence and presence of $^{39}$K atoms in the MOT.  Blue circles correspond to $^{40}$Ca$^+$ fluorescence data in the absence of the neutral atoms. Purple, red and orange circles show $^{40}$Ca$^+$ fluorescence when the ions are held together with $^{39}$K atoms with number densities of $6.8\times10^8$, $14.5\times10^8$, and $16.6\times10^8$, respectively. Dashed lines are exponential fits to the data.}
\end{figure}

The TOFMS signal, collected as a function of hold time, provides evidence of charge exchange, because our Paul trap holds  $^{40}$Ca$^+$ and $^{39}$K$^+$ ions equally well due to their comparable masses. That is, product $^{39}$K$^+$ ions are trapped and can be detected with TOFMS. Figure \ref{fig:TOF} shows the detected number of $^{40}$Ca$^+$ and $^{39}$K$^+$ ions as a function of hold time. The number of $^{40}$Ca$^+$ decreases with time with a simultaneous one-to-one increase in the number of $^{39}$K$^+$ ions.

Figure~\ref{fig:flr} shows typical $^{40}$Ca$^+$ fluorescence as a function of hold time in the absence and presence of $^{39}$K atoms. Without $^{39}$K atoms, the number of trapped, laser-cooled $^{40}$Ca$^+$ ions is constant for several minutes. When  $^{40}$Ca$^+$ ions are allowed to interact with  $^{39}$K atoms, however, the $^{40}$Ca$^+$ ions disappear in a few seconds as a result of charge exchange.  Decay rates measured by  $^{40}$Ca$^+$ fluorescence and TOFMS data are consistent. As TOFMS is destructive, both ion and atom traps need to be reloaded for each hold time. Hence, for most measurements described in this article $^{40}$Ca$^+$ ion fluorescence is used to measure decay rates. In fact, fluorescence has been the preferred measurement tool of ultra-cold chemical reaction rates \cite{Grier2009,Hall2011,Rellergert2011,Haze2015}.

\begin{figure}
\includegraphics[width=\columnwidth,trim=0 0 0 0,clip]{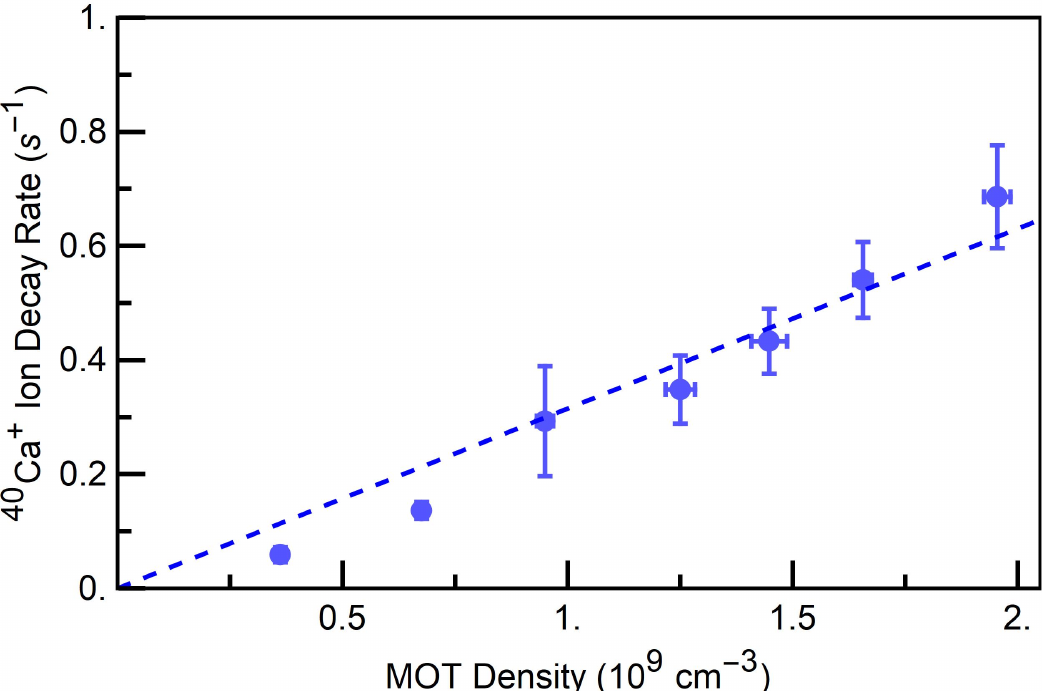}
\caption{\label{fig:fig5} Loss rate of $^{40}$Ca$^+$ ions in the presence of its 397-nm cooling light as a function of $^{39}$K number density in the MOT. A linear fit to the data assuming no charge exchange when there are no $^{39}$K atoms present gives a  rate coefficient of $(4.0\pm{0.2}) \times 10^{-10}$ cm$^3$/s. The detuning of the 397-nm laser  that cools $^{40}$Ca$^+$ is  such that the excited-state $^{40}$Ca$^+$($^2$P$_{1/2}$) population is 24 \%.}
\end{figure}

Figure \ref{fig:fig5} shows the rate of charge exchange as a function of $^{39}$K number density in a MOT when the fraction of $^{40}$Ca$^+$ ions in the excited $^2$P$_{1/2}$ state is 0.24. 
This excited-state fraction is controlled by the frequency detuning of the 397-nm 
$^{40}$Ca$^+$ cooling laser from the 4S to 4P$_{1/2}$ transition. The rate coefficient increases with $^{39}$K number density  and  a linear fit  assuming no charge exchange at zero $^{39}$K number density gives a rate coefficient of $(4.0\pm{0.2}) \times 10^{-10}$ cm$^3$/s. 

\begin{figure}
\includegraphics[width=\columnwidth,trim=0 0 0 0,clip]{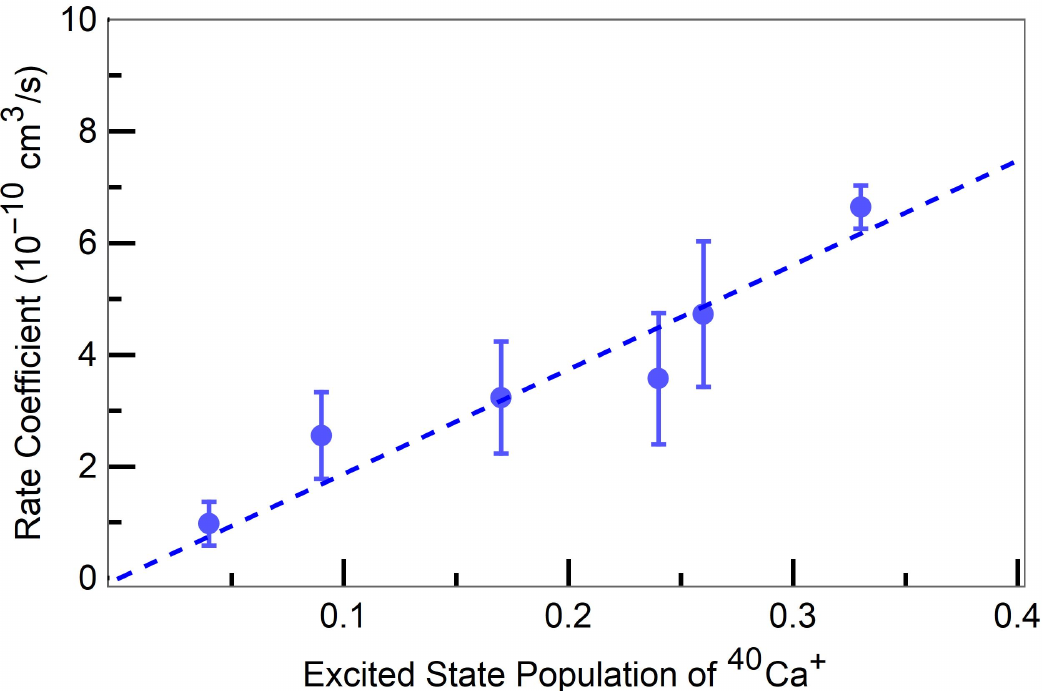}
\caption{\label{fig:population} Charge-exchange rate coefficient as a function of the $^{40}$Ca$^+$(4P$_{1/2}$) excited-state population. The error bars are one standard deviation uncertainties based on ten  measurements. 
}
\end{figure}

To examine the effect of the excited electronic state of $^{40}$Ca$^+$ ions on charge exchange, we varied the population in the 4P$_{1/2}$ state by adjusting the frequency detuning and intensity of the  $^{40}$Ca$^+$ lasers. Figure \ref{fig:population} shows these data. The rate coefficients increases linearly with excited-state population. We can elaborate that for the 100\% population of Ca$^+$ ions in the $^2P$ state the KCa$^+$ rate coefficient reaches value of $1.9\times 10^{-9}$ cm$^3$/s, which is only 0.55 times smaller than the Langevin rate for this system (K$_L$=$3.46 \times 10^{-9}$ cm$^3$/s) \cite{Langevin1905} . 
The rate coefficients of the charge exchange reaction for other systems such as LiCa$^+$ and LiYb$^+$ with Ca$^+$ and Yb$^+$ ions in an excited $^2P$ state have been reported in Refs.~\cite{Haze2015,Joger2017} to be $3.0\times10^{-10}$ cm$^3$/s (0.06 K$_L$) and $2.3\times10^{-9}$ cm$^3$/s  (0.5 K$_L$), respectively. In case of Yb$^+$ and Ba$^+$ colliding with Rb atoms, there are only measurements for the ions in the ground S state \cite{Sayfutyarova2013, Krukow2016}, the charge exchange rate coefficients for both mixtures are smaller than 10$^{-12}$ cm$^3$/s.

Finally, we conclude that when all $^{40}$Ca$^+$ ions are in the electronic ground state the charge-exchange rate coefficient is small, or more-precisely falls below our detection threshold.

\section{Theoretical models for the charge-exchange reaction}

\subsection{Charge exchange  in the presence of the  Ca$^+$ ion cooling laser}\label{ssec:ioncool}

Our experimental measurements provide clear evidence of the important role that
the excitation of   $^{40}$Ca$^+$ ions into the 4p$^2$P$_{1/2}$ state plays in changing the
charge exchange rate. To better understand the reaction mechanism we theoretically studied the excited-state
electronic potentials of the KCa$^+$ molecule in order to identify  processes that lead to high-yield reaction channels 
and the formation of K$^+$ ions as well as neutral excited-state Ca atoms.

\begin{figure}
		\includegraphics[width=0.5\textwidth,trim=0 5 0 0,clip]{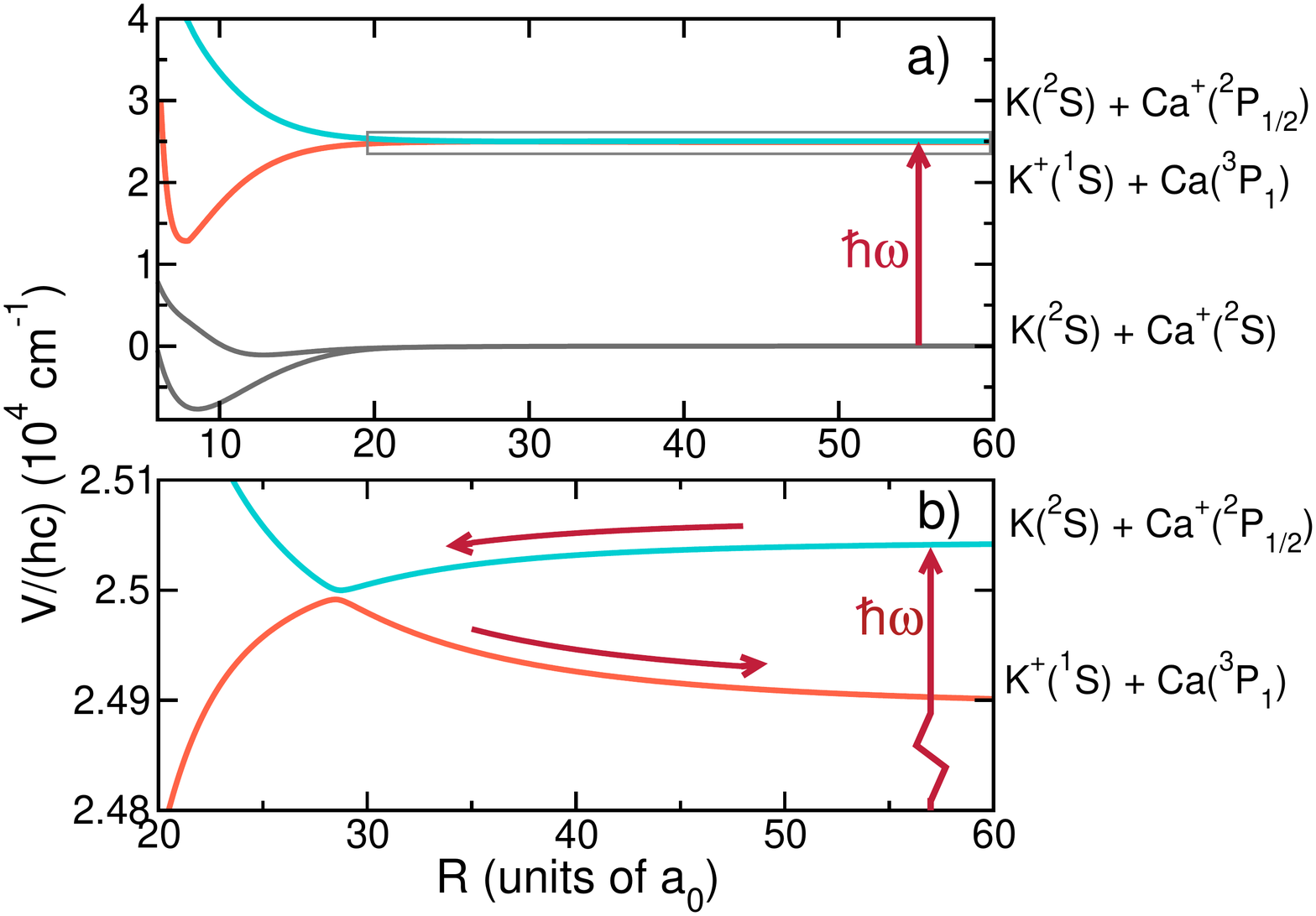}
\caption{a) Schematic of the relevant  electronic potential energy curves as functions of the K+Ca$^+$ separation $R$. Only curves relevant for charge-exchange starting from the absorption of a photon from
the ${\rm K}(^2{\rm S}_{1/2})+{\rm Ca}^+(^2{\rm S}_{1/2})$ to ${\rm K}(^2{\rm S}_{1/2})+{\rm Ca}^+(^2{\rm P}_{1/2})$ the channels are shown.  From low to high energy the two potentials dissociating to the ${\rm K}(^2{\rm S}_{1/2})+{\rm Ca}^+(^2{\rm S}_{1/2})$  limit
have $^1\Sigma^+$ and  $^3\Sigma^+$ symmetry, respectively.  The zero of energy is at the ${\rm K}(^2{\rm S}_{1/2})+{\rm Ca}^+(^2{\rm S}_{1/2})$ dissociation limit and $a_0$ is the Bohr radius.
Panel b) zooms in on the potentials leading to charge exchange  to ${\rm K}^+(^1{\rm S}_{1/2})+{\rm Ca}(^3{\rm P}_1)$
releasing only 150 cm$^{-1}$ equivalent kinetic energy. The gray box in panel a) corresponds to the energy and separation region shown in panel b).}
\label{fig:lr}  
\end{figure}

Enhanced collisional charge-exchange rate  occurs when molecular energy levels of Ca$^+$K and K$^+$Ca 
configurations are energetically close allowing for efficient charge exchange. As shown in Fig.~\ref{fig:lr} we find  
such conditions in collisional interactions between ground state K($^2$S$_{1/2}$) atoms and electronically-excited
Ca$^+$($^2{\rm P}_{j=1/2}$) ions on the one hand and between ground state K$^+$($^1$S$_0$) ions and metastable Ca(3d4p\,$^3{\rm P}_{j=1}$) atoms on the other. These two dissociation limits are only separated by 150 cm$^{-1}$ energy equivalent and have energies lying approximately ${E/hc=25000}$ cm$^{-1}$ above the  K(4s\,$^2$S$_{1/2}$) + Ca$^+$($^2$S$_{1/2}$) ground state. This latter energy corresponds to a photon with a wavelength of 397 nm, the wavelength of the cooling laser for the Ca$^+$ ion. Figure~\ref{fig:lr}b)  also shows that potentials dissociating to the two nearly-degenerate dissociation limits have an avoided crossing near $28 a_0$ leading to significant charge exchange. The zero of energy in Figure~\ref{fig:lr} is at the ${\rm K}(^2{\rm S}_{1/2})+{\rm Ca}^+(^2{\rm S}_{1/2})$ dissociation limit.

Figure~\ref{fig:lr} is based on an implicit approximation for our coordinate system that will be used throughout this paper. In charge exchange an electron moves from one atom to another, thereby, changing the center of charge and mass of the dimer. We assume that the effects of these changes on rate coefficients is negligible and express all molecular interactions in terms of separation $\vec R=(R,\theta,\varphi)$ in spherical coordinates defined as the distance $R$ between and orientation $\theta,\varphi$ of the center of mass of neutral $^{39}$K with respect to that of the center of mass of ionic $^{40}$Ca$^+$ in a laboratory-fixed coordinate frame. The reduced mass  $\mu=m_{\rm K}m_{\rm Ca}/(m_{\rm K}+m_{\rm Ca})$ is  that of the $^{39}$K+$^{40}$Ca$^+$ system, where $m_{\rm K}$ and $m_{\rm Ca}$ are
the masses of $^{39}$K and $^{40}$Ca$^+$, respectively.

The reaction between atoms and ions is governed by four interactions: a) an isotropic and anisotropic  $C_4/R^4$ induction interaction between the charge of an ion and the induced dipole moment of a neutral atom; b) an anisotropic $C_3/R^3$ interaction between the ion charge and the quadrupole moment of an excited neutral atom; c)  the spin-orbit interaction that defines splitting of the Ca$^+$(4p\,$^2{\rm P}_{j}$) and Ca(3d4p\,$^3{\rm P}_{j}$ fine-structure states; d) the molecular rotation. Both $C_4$ and $C_3$ coefficients orientationally dependent and defined byproperties of the neutral atom. Dispersive van-der-Waals potentials do not significantly contribute to the atom-ion interaction for $R>20a_0$. 

If we ignore the rotation of the molecule for the moment, a convenient basis  for the reaction from
${\rm K}({\rm 4s}\, ^2{\rm S}_{j_a})+{\rm Ca}^+({\rm 4p}\, ^2{\rm P}_{j_b})$ to
${\rm K}^+(^1{\rm S}_{j'_a})+{\rm Ca}({\rm 3d4p}\,^3{\rm P}_{j'_b})$
is the body-fixed product basis 
\begin{equation}
 |q_a, j_a,\Omega_a\rangle |q_b,j_b,\Omega_b\rangle\,,
 \label{eq:basisa}
 \end{equation}
where $q_i=0,+1$ is charge state of the atom or ion $i=a$ and $b$ for K/K$^+$ and Ca/Ca$^+$, respectively,
and $\Omega_i$ is the  projection quantum number of  the total atomic/ionic angular momentum $\vec\jmath_i$
along the interatomic axis $\vec R$. We  use that within the context of this charge-exchange process
labels $q_i$, $j_i$, and $\Omega_i$ uniquely specify an atomic/ionic state.
A list of states can be found in Table~\ref{tab:c4}.

\begin{table}
	\caption{The quantum numbers for our body-fixed basis states for the ${\rm K}({\rm 4s}\, ^2{\rm S}_{j_a})+{\rm Ca}^+({\rm 4p}\, ^2{\rm P}_{j_b})$ to
${\rm K}^+(^1{\rm S}_{j'_a})+{\rm Ca}({\rm 3d4p}\,^3{\rm P}_{j'_b})$ reaction
 and the corresponding diagonal $C_3$ and $C_4$ coefficients in atomic units. Channels are uniquely described by the charge $q_i$, atomic angular
		momenta $j_i$, and its body-fixed projection $\Omega_i$ on the
		internuclear axis with $i=a$ and $b$ for K and Ca, respectively.
		Potentials are degenerate for $-\Omega$ and $\Omega$, where
		$\Omega=\Omega_a+\Omega_b$. }
		\label{tab:c4}
	\begin{ruledtabular}
\begin{tabular}{cccccccc}
	\multicolumn{3}{c}{K or K$^+$}  &  \multicolumn{3}{c}{Ca or Ca$^+$}  \\
	\cline{1-3}
	\cline{4-6}
	$q_a$ & $j_a$ & $\Omega_a$ & 
	$q_b$ & $j_b$ & $\Omega_b$ & $C_3$ & $C_4$\\
	\hline
	0  & $1/2$ & $1/2$ &+1 & 1/2 & $-1/2$ &      0 & $-$145.29 \\
	0  & $1/2$ & $-1/2$ &+1 & 1/2 & 1/2 &      0 & $-$145.29 \\
	0  & 1/2 & 1/2 &  +1 & 1/2 & 1/2 &      0 & $-$145.29 \\
	+1 &   0 &   0 &   0 &   0 &   0 &      0 & $-$641.0  \\
	+1 &   0 &   0 &   0 &   1 &   0 &  5.874 &   97.95  \\
	+1 &   0 &   0 &   0 &   1 &   1 & $-$2.937 & $-$1015.5  \\
	+1 &   0 &   0 &   0 &   2 &   0 & $-$5.874 & $-$1404.7  \\
    +1 &   0 &   0 &   0 &   2 &   1 & $-$2.937 & $-$1027.9  \\
    +1 &   0 &   0 &   0 &   2 &   2 &  5.874 &   102.7  \\
\end{tabular}
\end{ruledtabular}
\end{table}

The matrix elements of the two long-range molecular interactions 
are diagonal in and independent of the quantum numbers of the ion and
only depend on the state of the neutral atom. In fact,
\begin{eqnarray*}
\frac{\langle 0, j_s,\Omega_s|C_{3}|0,j'_s,\Omega'_s\rangle}{R^3}
 &=&\delta_{\Omega_s\Omega'_s}  \frac{q}{R^3} (-1)^{j_s-\Omega_s}\\
 && \quad \times
\begin{pmatrix}
       j_s & 2 & j'_s \\
      -\Omega_s & 0 & \Omega_s
       \end{pmatrix}
       \langle j_s ||Q_2|| j'_s \rangle  \nonumber
\end{eqnarray*}
 in atomic units for the neutral atom with quantum numbers $j_s\Omega_s$  \cite{Tomza2019}.
Here, $q=+1$ for the corresponding ion, $\delta_{\Omega\Omega'}$ is the Kronecker delta, and (:::) denotes a Wigner 3-$j$ symbol. 
The  matrix elements of the induction potential for  K(4s\,$^2$S) are
\[
\frac{\langle 0, j_s,\Omega_s|C_{4}|0,j'_s,\Omega'_s\rangle}{R^4}=-
\delta_{j_s\Omega_s,j'_s\Omega'_s}
\frac{q^2}{2R^4}\alpha_{0,s}
   \,,
\]
 while the diagonal matrix elements of the induction potential for Ca(3d4p\,$^2$P) states are
\begin{eqnarray*}
\frac{\langle 0, j_s,\Omega_s|C_{4}|0,j_s,\Omega_s\rangle}{R^4} 
&=& -\frac{q^2}{2R^4} \quad \\
 \times &&\left[\alpha_{0,s}
  +\alpha_{2,s}\frac{3\Omega_s^2-j_s(j_s+1)}{j_s(2j_s-1)}\right]\nonumber \,,
\end{eqnarray*}
from Ref.~\cite{Mitroy2010}. Off-diagonal
matrix elements, coupling states with $|\Delta j_s|\le 2$, can also be found in Ref.~\cite{Mitroy2010}.
Finally,  $\langle j_s ||Q_2|| j_s \rangle$, $\alpha_{0,s}$,  and $\alpha_{2,s}$
are the reduced quadrupole-moment matrix elements and  static scalar and tensor polarizabilities of the neutral atom $s$ with angular momentum $j_s$, respectively. 
The sum  $\Omega=\Omega_a+\Omega_b$ is  conserved for both interactions.

We have calculated the reduced matrix element of the quadrupole moment of  Ca(3d4p\,$^3$P) states using a non-relativistic multi-configuration-interaction method with an all-electron correlation-consistent polarized valence-only quintuple zeta (CC-pV5Z) basis set \cite{Koput2002}.  For the triplet states calculations have been performed in the $B_{3u}$ representation. More details of our calculation can be found Supplemental Material.
This calculation provides the expectation value of the quadrupole operator $\hat{Q}_{20}$ for state $| 0, j_s=1,\Omega_s=0\rangle$ of neutral Ca(3d4p\,$^3$P). The value is $-11.748$ a.u. with an one-standard-deviation
uncertainty of 3.0 a.u. (Here, a.u. is an abbreviation for atomic unit and we keep all digits to avoid roundoff problems.)
Of course, the quadrupole moment for an S state is  zero.
 
We used the static  scalar polarizability of $290.48$ a.u.  for K($4{\rm s}\,^2{\rm S}_{1/2}$) from Ref.~\cite{Holmgren2010}.  For  Ca(3d4p\,$^3{\rm P}_{0,1,2}$) states we have computed $\alpha_{0,s}$ and $\alpha_{2,s}$ using
transition dipole moments from \cite{Yu2018} and perturbation theory.
Values for quadrupole moments and static polarizabilities are summarized in Table~\ref{tab:c3}.
Table \ref{tab:c4} lists the diagonal $C_3$ and $C_4$ coefficients for the relevant  channels $ |q_a, j_a\Omega_a\rangle |q_b,j_b\Omega_b\rangle$.

The charge-exchange-inducing coupling between the body-fixed channels is computed based on the application of the  Heitler-London method, in which the electron in the 4s orbital of K moves to the unoccupied 3d orbital
of Ca$^+$. The coupling is then given by the overlap integral of these non-relativistic atomic Hartree-Fock  orbitals and the electron-nucleus Coulomb interaction potential. The coupling matrix element between  channels  $|0, j_a\Omega_a\rangle |{+1},j_b\Omega_b\rangle$ and  $|{+1}, j'_a\Omega'_a\rangle |0,j'_b\Omega'_b\rangle$ with the same $\Omega=\Omega_a+\Omega_b=\Omega'_a+\Omega'_b$=0, 1  are $\sqrt{\frac{2}{30}}O(R)$ and $\sqrt{\frac{1}{30}}O(R)$, respectively, where
\begin{eqnarray}
 O(R)&=&  \int d^3r \,\phi_{\rm K,4s}(\vec r-\vec r_{\rm K}) \frac{e^2}{4\pi\epsilon_0}
 \left[ \frac{1}{|\vec r-{\vec r}_{\rm K}|} + \frac{1}{|\vec r-{\vec r}_{\rm Ca}|}\right] 
 \nonumber \\
  && \quad\quad \quad\quad \times\phi_{\rm Ca, 3d}(\vec r-\vec r_{\rm Ca})\,,
\end{eqnarray} 
 $\phi_{\rm K,4s}(\vec r-\vec r_{\rm K})$ and $\phi_{\rm Ca, 3d}(\vec r-\vec r_{\rm Ca})$
are the spatial-components of the K(4s) and Ca(3d) Hartree-Fock electron orbitals, respectively, and $\vec r_{i}$  
is the location of  nucleus $i= {\rm K}$ or Ca. For our choice of coordinate system $R=|\vec r_{\rm K}-\vec r_{\rm Ca}|$ to good approximation. The function $O(R)$ is $hc\times 5.6$ cm$^{-1}$ at $R=28a_0$ and rapidly goes to zero for $R\to\infty$.

\begin{table}
	\caption{Atomic parameters in atomic units used to generate the interaction potentials. See text for definitions.}\label{tab:c3}
\begin{tabular}{lcr@{.}lr@{.}lr@{.}l}
\hline\hline
\multicolumn{2}{c}{Term} & \multicolumn{2}{c}{$\alpha_0$} & \multicolumn{2}{c}{$\alpha_2$} & \multicolumn{2}{c}{$\langle j || Q_2 || j\rangle$}  \\
\hline
K (4s) &$^2{\rm S}_{1/2}$ & 290 & 58  \\
Ca (4s$^2$) &$^1{\rm S}_0$ & 157&1    \\
Ca (3d4p) &$^3{\rm P}_0$ & 1282&0   \\
Ca (3d4p) &$^3{\rm P}_1$ & 1288&7 & 742&3 & $-$16 &086 \\
Ca (3d4p) &$^3{\rm P}_2$ & 1302&0 & $-$1507&4 & 32 &173 \\
\hline\hline
\end{tabular}
\end{table}

\begin{figure*}
		\includegraphics[width=\textwidth,trim=0 0 0 0,clip]{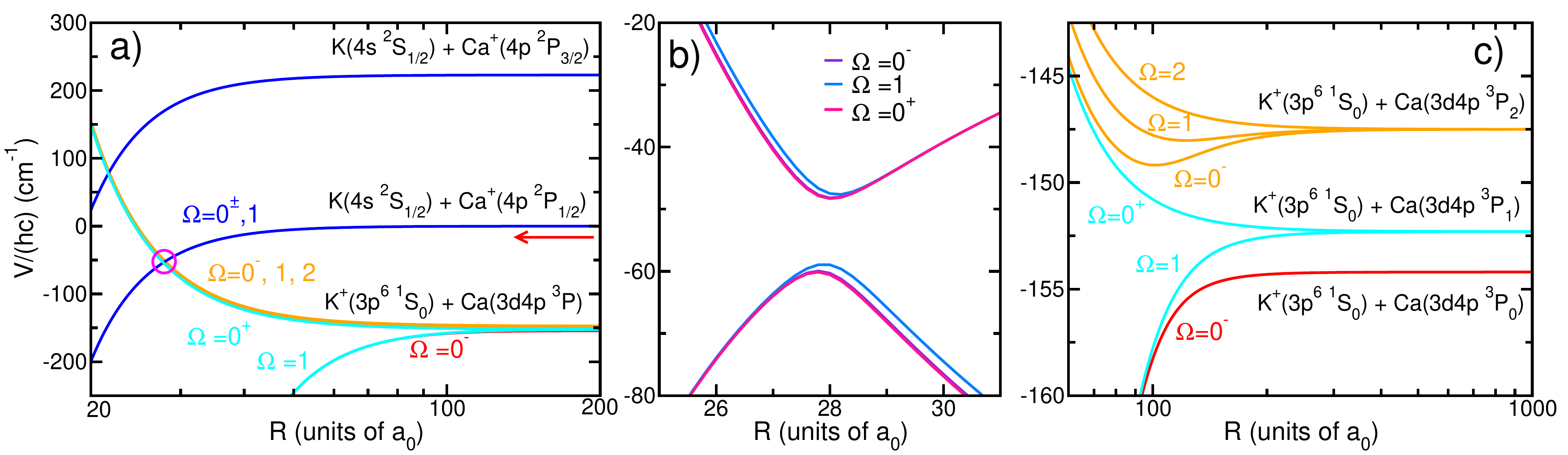}
\caption{Long-range adiabatic potential energy curves with $\Omega=0^\pm$, 1, and 2 as functions of the atom-ion separation $R$.
Panels a) and c) have been computed without the charge-exchange-inducing coupling. 
Panel (b) shows  avoided crossings among  adiabatic potentials due to the charge-exchange-inducing coupling near
$R=28a_0$. The zero of energy in all panels is located at the K(4s $^2{\rm S}_{1/2}$) + Ca$^+$(4p $^2{\rm P}_{1/2}$)
entrance channel (red arrow in panel a).
\label{fig:lr2}  
}
\end{figure*}

Figure~\ref{fig:lr2} shows three views of the  long-range adiabatic $\Omega = 0^{\pm}$, 1 and 2 potentials for the excited-state charge-exchange reaction.  
Panels a) and c) have been found by diagonalizing the potential matrix at each $R$ in the body-fixed basis of 
Eq.~\ref{eq:basisa}  without the charge-exchange-inducing coupling, while in panel b) this coupling has 
been included.
Consequently, curves dissociating to the K(4s $^2{\rm S}_{1/2}$) + Ca$^+$(4p $^2{\rm P}$)  limits in Fig.~\ref{fig:lr2}a) cross those dissociating to the K$^+$($^1{\rm S}_0$) + Ca(3d4p $^3{\rm P}$) limits.
These crossings become avoided crossings once charge-exchange-inducing couplings are included,
as shown in Fig.~\ref{fig:lr2}b) around $R = 28 a_0$.
The (avoided) crossings for the curves dissociating to the K(4s $^2{\rm S}_{1/2}$) + Ca$^+$(4p $^2{\rm P}_{3/2}$)
limits are not energetically-accessible from our K(4s $^2{\rm S}_{1/2}$) + Ca$^+$(4p $^2{\rm P}_{1/2}$) 
entrance channels and need not be considered further.
Charge exchange can only occur between states with the same $\Omega$, which in our system only occurs between $\Omega = 0^{\pm}$ and 1 potentials.  
The asymptotic splittings between the entrance channel and the  exit channels is approximately $hc \times 150$ cm$^{-1}$. The zero of energy in all panels is located at the K(4s $^2{\rm S}_{1/2}$) + Ca$^+$(4p $^2{\rm P}_{1/2}$)
entrance channel.

For the experimental temperature near 0.1 K and a long-range induction potential,
the number of partial waves or relative orbital angular momenta of the atom-ion system, $\vec \ell$, contributing to 
the charge-exchange rate coefficient is large. Then we are justified in approximating the rotational Hamiltonian 
of the diatom and use the infinite-order sudden approximation~\cite{Nikitin2006}.
This corresponds to performing coupled-channels calculations for each triple $J$, $M$, and $\Omega$ in the 
channel basis $\sqrt{(2J+1)/8\pi^2}D^{J*}_{M\Omega}(\varphi,\theta,0))|q_a,j_a \Omega_a\rangle | q_b,j_b \Omega_b \rangle$  
ignoring couplings between different $\Omega^\pm$ and using a diagonal centrifugal potential with matrix element 
$\hbar^2J(J+1)/2\mu R^2$ for each channel  independent of $M$ and $\Omega^\pm$.
Here, $D^j_{mm'}(\alpha\beta\gamma)$ is a Wigner rotation matrix, $\Omega=\Omega_a+\Omega_b$, 
$\vec J=\vec \jmath_a+\vec \jmath_b+\vec\ell$ is the total molecular angular momentum with projection quantum 
number $M$. For $\Omega=0^-$, $0^+$, and 1 channels are coupled for each 
$J$ and $M$, respectively, as we can ignore the effects of the energetically-closed K(4s $^2{\rm S}_{1/2}$) + Ca$^+$(4p $^2{\rm P}_{3/2}$) channels. For a valid coupled-channels calculation we extend the potentials to $R\to 0^+\, a_0$ to have
a repulsive inner wall  by adding repulsive $C_{12}/R^{12}$ potentials to the diagonal potential matrix elements in the channel states. Typical examples can be seen in Fig.~\ref{fig:lr}.)

We propagate the wavefunctions in the coupled-channels calculations for each $J,M,\Omega^\pm$ using 
Gordon's propagator~\cite{Gordon1969} and match to scattering boundary conditions at large $R$ to obtain 
 partial charge-exchange rate coefficients $K_{JM,\Omega^\pm}(E)$, where $E$ is the entrance-channel collision energy and have summed over all exit channels. 
\begin{equation}
K(E) = \sum_{\Omega^p=0^+,0^-,1} g_{\Omega^p}\sum_{J=0}^\infty (2J+1) K_{JM,\Omega^p}(E) \;,
\label{eq:K}
\end{equation}
where $g_{0^+}$=1/4,  $g_{0^-}$= 1/4, and $g_{1}$=1/2. 

\begin{figure}
	\includegraphics[width=\columnwidth,trim=10 30 60 70,clip]{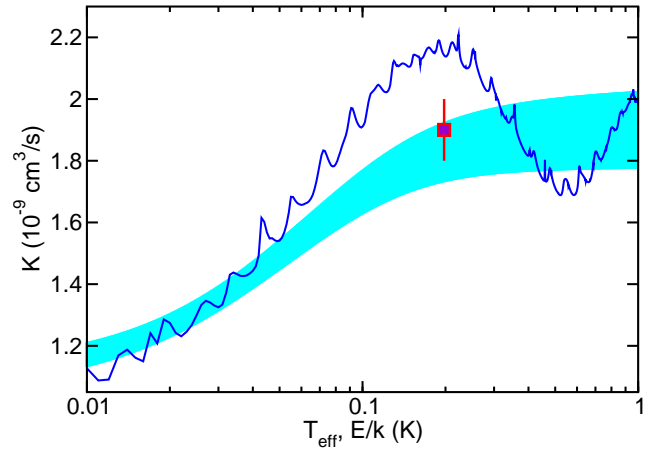}
	\caption{
	${\rm K}+{\rm Ca}^+\to{\rm K}^++{\rm Ca}$ charge-exchange  rate coefficients in the presence of the 
	 Ca$^+$-ion cooling laser based on coupled-channels calculations as a function of collision energy $E$ or 
	 effective temperature $T_{\rm eff}$ as defined in the text.
	The brown curve shows an example of  rate coefficient $K(E)$ as a function of collision energy $E$.
	The cyan colored band corresponds to the
	thermally-averaged rate coefficient as a function of $T_{\rm eff}$.
	The width of the band reflects our estimate of the one-standard deviation uncertainty of the coupled channels
	calculations. Finally,  the red marker with one-standard deviation error bars 
	is our experimentally measured thermally-averaged rate coefficient at $T_{\rm eff}=0.2$ K.}
	\label{fig:RCT}
\end{figure}

Finally, we thermally average assuming a Gaussian distribution with a temperature $T_{\rm K}$ and
$T_{\rm Ca}$ for $^{39}$K and $^{40}$Ca$^+$, respectively. This distribution is used for the trapped ion cloud and 
based on the evidence that the effective trapping area is limited to a certain radius \cite{Delahays2019}, which increases the likely-hood of low energy ions above that of a Maxwell-Boltzmann distribution. The thermalized total rate coefficient then only depends on the effective temperature
\[
    T_{\rm eff} =  \frac{m_{\rm K} T_{\rm Ca} +m_{\rm Ca}T_{\rm K}}{M}  \approx \frac{m_{\rm K} }{M} T_{\rm Ca}
     \,.
\]
where $M = m_{\rm K} + m_{\rm Ca}$ and the second approximate equality follows from $T_{\rm K}\ll T_{\rm Ca}$ in our experiments.
Figure \ref{fig:RCT} shows  results for the  total charge-exchange rate coefficient from our coupled-channels calculations and compares them to our  measurement at $T_{\rm eff} =0.2$ K. The figure first shows a rate coefficient as a function of collision energy $E$.
The rate coefficient has many peaks resulting  from shape resonances contained by the barrier created by the attractive
induction potential and the repulsive centrifugal potential. At $E/k<0.2$ K already twelve such resonances can be observed.
Here, $k$ is the Boltzmann constant.
In fact,  for $E/k=0.2$ K at least forty $J$ contribute significantly to the total charge-exchange rate coefficient.
Once the rate coefficient is thermally-averaged the peaks smooth out and the rate coefficient  is a slowly increasing function with $T_{\rm eff}$ for $T_{\rm eff}<1$ K.
Finally, we have studied the effects of changing the  quadrupole moment and thus the $C_3$ coefficient within
its 20\% uncertainty on the thermalized rate coefficient. As shown in Fig.~\ref{fig:RCT} we observe that the  rate coefficient
changes by 10\% to 20\% for temperatures between 0.01 K and 1 K. The theoretical rate coefficient is in agreement
with the experimental value at $T_{\rm eff} =0.2$ K.

\subsection{Other charge-exchange pathways in the K and Ca$^+$ system}\label{ssec:other}

In this subsection we discuss three other processes that lead to charge exchange between colliding a K atom and a 
Ca$^+$ ion. Their rate coefficients turn out to be much smaller than that for the process described in the previous section.
A priori this was not obvious. For completeness, we briefly describe our calculations of these processes. The first process (pathway I) is simply given by
the non-radiative process
\[
{\rm K}(^2{\rm S}_{1/2})+{\rm Ca}^+(^2{\rm S}_{1/2})  \to {\rm K}^+(^1{\rm S}_{0})+
{\rm Ca}(^1{\rm S}_0) \,.
\] 
As already schematically shown in Fig.~\ref{fig:lr}a, K+Ca$^+$  collide on a  
$^{1}\Sigma^+$ and $^{3}\Sigma^+$ potential of the KCa$^+$ molecule. 
A second process (pathway II) is a radiative process and combines
\[
{\rm K}(^2{\rm S}_{1/2})+{\rm Ca}^+(^2{\rm S}_{1/2})  \to {\rm K}^+(^1{\rm S}_{0})+
{\rm Ca}(^1{\rm S}_0) + \hbar\omega
\]
and
\[{\rm K}(^2{\rm S}_{1/2})+{\rm Ca}^+(^2{\rm S}_{1/2})
\to{\rm KCa}^+({\rm X}\,^1\Sigma^+) + \hbar\omega 
\,,
\] 
where the colliding atom and ion spontaneously emit photons with energy $\hbar\omega$. In the second contribution
a molecular ion is formed.
Finally, non-radiative pathway III is 
\[
{\rm K}(^2{\rm P}_{3/2})+{\rm Ca}^+(^2{\rm S}_{1/2})  \to {\rm K}^+(^1{\rm S}_{0})+{\rm Ca}(^1{\rm P}_1)\,,
\]
enabled by the excitation of  potassium atoms  to the 4p\,$^2$P$_{3/2}$ state due to the absorption of a MOT photon. 

Unlike for the calculations in the previous subsection, we have calculated accurate Born-Oppenheimer potential energy curves $V_s(R)$ for all separations $R$.
Since the structure of KCa$^+$ was  unknown, we calculated  all $^{2S+1}\Lambda^\pm$ non-relativistic  potentials  dissociating to the eight energetically-lowest  
asymptotes of K+Ca$^+$ and K$^+$+Ca, using a  multi-reference configuration-interaction (MRCI) method within the MOLPRO program package~\cite{Werner2012}. 
In addition, we  required the non-adiabatic coupling
function $Q_{s,s'}(R)=\langle \psi_s;R|d/dR|\psi_{s'};R\rangle$  and electronic transition dipole moment $d_{ss',\rm e}(R)=\langle \psi_s;R|\hat{d}|\psi_{s'};R\rangle $ between the two energetically-lowest $^{1}\Sigma^{+}$ states,
where kets $|\psi_{s};R\rangle $ denote the $R$-dependent adiabatic electronic eigen states and
$\hat{d}$ is the electronic dipole operator.
Details of these calculations are given in Supplemental Material. 

\begin{figure}
	\includegraphics[width=0.9\columnwidth,trim=0 0 0 0,clip]{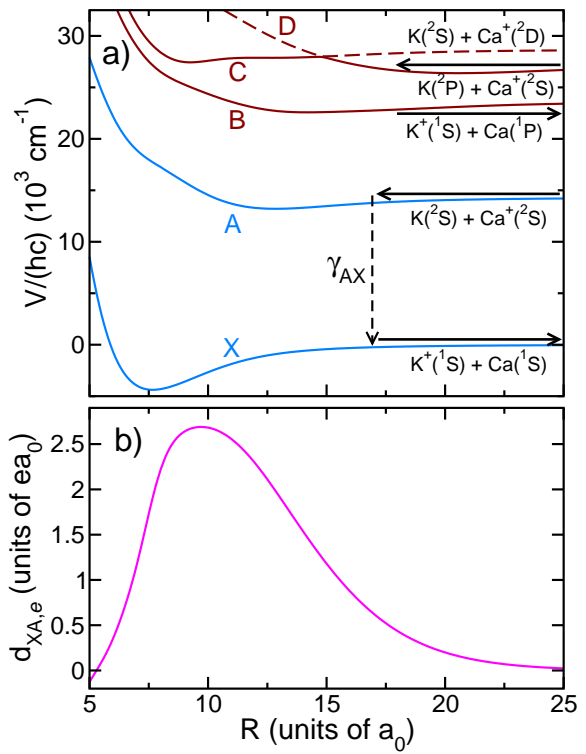}
	\caption{Panel a) The energetically-lowest five  $^1\Sigma^+$ Born-Oppenheimer potential energy curves, labeled X, A, B, C, and D, as functions of separation $R$. The potentials have been computed with a multi-reference configuration-interaction method. The arrows indicate the entrance channels for the three pathways discussed in Sec.~\ref{ssec:other}. The blue  curves are relevant for pathways I and III. Red curves are relevant for pathway II.  The vertical dashed
arrow labeled $\gamma_{\rm AX}$ represents spontaneous emission from state A to X in pathway III. The zero of energy is at the ${\rm K}^+(^1{\rm S}_{0})+{\rm Ca}(^1{\rm S}_0) $ limit.
Panel b) Electronic transition dipole moment $d_{{\rm XA},e}(R)$ between the
ground X$^1\Sigma^+$ and  excited A$^1\Sigma^+$ adiabatic Born-Oppenheimer 
states from our MRCI calculations. The  potentials and dipole moment for $R<5 a_0$
are not shown because the entrance channel A$^1\Sigma^+$ potential is 
repulsive there and not relevant for the calculation of the rate coefficient.}
\label{fig:other}
\end{figure}

Figure \ref{fig:other}a) shows the five $^1\Sigma^+$ Born-Oppenheimer potentials $V_s(R)$ relevant for the processes described in this subsection.
The long-range interaction between an atom in an S-state and an ion in an S-state is isotropic, has $^1\Sigma^+$ symmetry, and is described by $-C_{4}/R^{4}-C_{6}/R^{6}$. For the energetically-lowest X$^1\Sigma^+$ state $C_{4}= 78.5 E_{\rm h} a_0^4$~\cite{Porsev2006}, while for the first-excited A$^1\Sigma^+$ state  $C_{4}= 145.29 E_{\rm h} a_0^4$~\cite{Holmgren2010} and $C_{6}= 3866 E_{\rm h} a_0^6$~\cite{Kaur2015}. Here, $E_{\rm h}$ is the Hartree energy and $a_0$ is the Bohr radius

\vspace*{3mm}
\noindent
{\bf Non-radiative charge exchange by Pathway I}

The description of cold charge-exchange collisions  by pathway I involves the energetically-lowest two $^1\Sigma^+$ Born-Oppenheimer 
potentials, one $^3\Sigma^+$ potential, the non-adiabatic coupling between the two $^1\Sigma^+$ potential leading to 
charge exchange, but also the atomic hyperfine interaction of 
$^{39}{\rm K}(4{\rm s}\,^2{\rm S}_{1/2})$ between its electron and nuclear spin. The nuclear spin of $^{40}$Ca$^+$ is zero.
For obtaining a qualitative estimate of rate coefficients for this pathway we can ignore the atomic hyperfine interaction Hamiltonian and then only account for the spin degeneracies.

With these approximations we can set up coupled-channels equations for charge exchange that includes only the isotropic 
X$^1\Sigma^+$ and A$^1\Sigma^+$ Born-Oppenheimer potentials, which are separated in energy by more than $hc\times10^3$ cm$^3$, using a two-channel diabatic representation that 
 can then be used to solve the coupled-channels equations~\cite{Nikitin2006}.
That is, we start with the $R$-dependent electronic eigenstates $|\psi_{s};R\rangle $ with $s$=X or A for the two corresponding adiabatic 
Born-Oppenheimer states, respectively, and construct $R$-independent diabatic electronic basis functions $|\phi_{i}\rangle $ with $i=1,2$ 
based on the unitary transformation
\begin{equation}
 \left(\begin{array}{c}
 |\phi_1\rangle \\
 |\phi_2\rangle
 \end{array}\right)
 =\left(
 \begin{array}{cc}
 \cos \theta(R) & -\sin\theta(R) \\
 \sin\theta(R) & \cos\theta(R)
 \end{array}\right)\left(\begin{array}{c}
 |\psi_{\rm X};R\rangle \\
 |\psi_{\rm A};R\rangle
 \end{array}\right)   
\end{equation}
with mixing angle 
$\theta(R) = \int_{R}^{\infty}Q_{\rm X,A}(R')dR'$ and non-adiabatic coupling $Q_{\rm X,A}(R)$ as obtained
by our MRCI calculations.
The potential matrix elements $U_{ij}(R)$ in the diabatic basis $|\phi_{i}\rangle $ are then
\begin{eqnarray}
&&U_{11}(R) =  V_{\rm X}(R)\cos^2 \theta +  V_{\rm A}(R)\sin^2 \theta \,,\nonumber\\
&&U_{22} (R)=  V_{\rm X}(R)\sin^2 \theta +  V_{\rm A}(R)\cos^2 \theta  \,, \\
&&U_{12} (R)=  U_{21}(R) = (V_{\rm A}(R) - V_{\rm X}(R))\cos \theta \sin \theta \nonumber \,.
\end{eqnarray}

\begin{figure}
 \includegraphics[width=1\columnwidth,trim=0 0 0 0,clip]{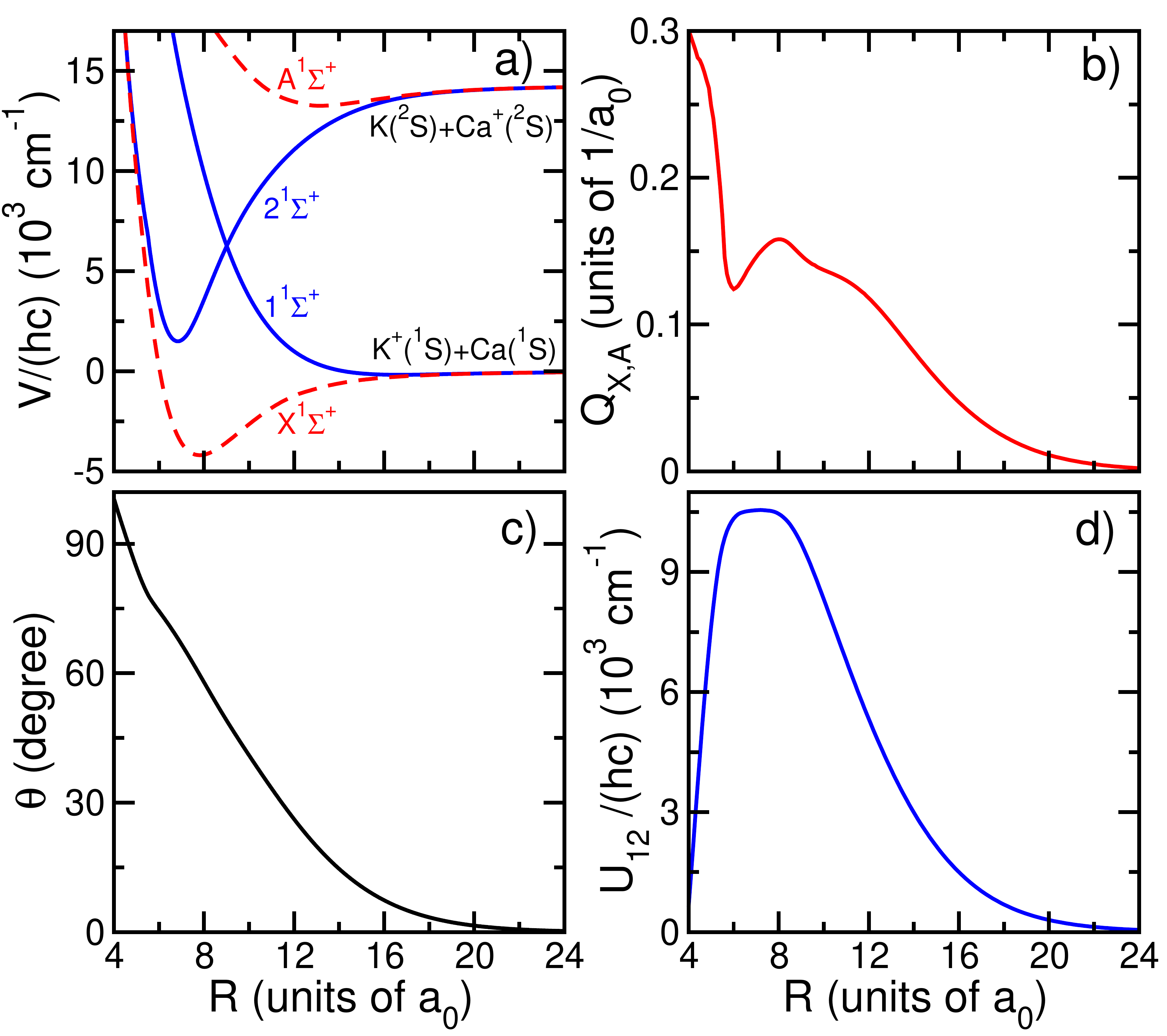}
 \caption{Diabatic representation of the ${\rm K}(^2{\rm S}_{1/2})+{\rm Ca}^+(^2{\rm S}_{1/2})  \to {\rm K}^+(^1{\rm S}_{0})+
{\rm Ca}(^1{\rm S}_0)$ charge-exchange process.
 (a) The ground X$^1\Sigma^+$ and  excited A$^1\Sigma^+$  adiabatic Born-Oppenheimer
 potentials, $V_{\rm X}(R)$ and $V_{\rm A}(R)$, (dashed red lines)
 and 1$^1\Sigma^+$ and 2$^1\Sigma^+$ diabatic potentials,
 $U_{11}(R)$ and $U_{22}(R)$, (solid blue lines) as functions of separation $R$.
 (b) the non-adiabatic coupling function $Q_{\rm X,A}(R)$ between the two 
 $^{1}\Sigma^{+}$ adiabatic Born-Oppenheimer potentials as a function of $R$. 
 Finally, panels c) and d) show  mixing angle $\theta(R)$ and  diabatic coupling function $U_{12}(R)$ as 
 functions of  separation $R$, respectively.}
\label{diabatic}
\end{figure}

Figures \ref{fig:other}a) and \ref{diabatic} shows all ingredients and results of the
process to create $R$-independent electronic wavefunctions.
Specifically, the inputs are $V_{\rm X}(R)$ and $V_{\rm A}(R)$
in Fig.~\ref{diabatic}a) and $Q_{\rm X,A}(R)$ in Fig.~\ref{diabatic}b).
The zero of energy in Figures \ref{fig:other}a) is at the ${\rm K}^+(^1{\rm S}_{0})+{\rm Ca}(^1{\rm S}_0) $ limit.
The resulting mixing angle is shown in Fig.~\ref{diabatic}c).
Finally, diabatic potentials $U_{ii}(R)$ and  coupling $U_{12}(R)$ are shown in 
Fig.~\ref{diabatic}a) and d), respectively. We note that the diabatic and adiabatic potentials in Fig.~\ref{diabatic}a) have very different shapes. This difference is due to the fact that the original adiabatic potentials are energetically well separated. The diabatic coupling near the $R=9 a_0$ crossing point is very large. Conventionally, the diabatization is only performed for the adiabatic potentials that are energetically close. 

As the diabatic potential matrix elements are isotropic, we can use  basis functions 
$|\phi_i \rangle Y_{\ell m}(\theta,\varphi)$ with $i$=1 and 2
to expand the total molecular wavefunction. Here, $Y_{\ell m}(\theta,\varphi)$ is a spherical harmonic,
$\ell$ is the partial wave
or the rotational angular momentum quantum number, and $m$ is its
projection quantum number in the laboratory frame. 
We then solve two-channel radial coupled-channels equations for each partial wave $\ell$
and $m=0$ by numerically propagation and obtain the partial non-radiative charge-exchange (nRCE)
rate coefficients
$K_{\ell m}^{\rm nRCE}(E)$. The rate coefficient is the same for each $m$
and $E$ is the entrance-channel collision energy. The total rate coefficient is then
\begin{equation}
K_{\rm nRCE}(E) = v \sigma_{\rm nRCE}(E)=g\sum_{\ell=0}^\infty (2\ell+1) 
   K_{\ell 0}^{\rm nRCE}(E) \,,
\end{equation}
where  $v=\sqrt{2E/\mu}$ is the relative velocity and ${g=1/4}$ is a statistical probability that accounts for the fact that  the entrance threshold 
K$(^{2}{\rm S}_{1/2})+$Ca$^{+}(^{2}{\rm S}_{1/2})$ has  singlet and  triplet total electron spin channels
and that only the singlet $^1\Sigma^+$ channel leads to charge exchange.  

\begin{figure}
 \includegraphics[width=\columnwidth,trim=0 0 0 0,clip]{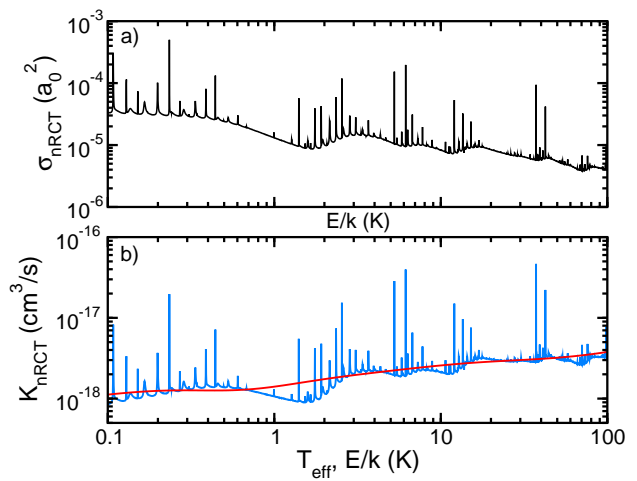}
 \caption{ Panel a) Non-radiative charge-exchange cross section from
 K$(^{2}{\rm S}_{1/2})+$Ca$^{+}(^{2}{\rm S}_{1/2})$ to K$^{+}(^{1}{\rm S}_0)+$Ca$(^{1}{\rm S}_0)$ 
 as a function of the entrance-channel collision energy.  Panel b) Thermalized (red line) and non-thermalized rate coefficients as a function of temperature and collision energy based on the cross section shown in panel a).}
\label{rateXA}
\end{figure}

Figure~\ref{rateXA}a) shows our non-radiative charge-exchange cross section $\sigma_{\rm nRCE}(E)$ for pathway I as a function of collision energies between 
$k\times 0.1$ K and $k\times 100$ K. 
The sharp resonances are again due to shape resonances of the induction potential  from the forty partial
waves that contribute significantly to $\sigma_{\rm nRCE}(E)$ at $E/k=0.2$ K.
Thermally averaging as in Sec.~\ref{ssec:ioncool} assuming temperatures $T_{\rm K}$ and $T_{\rm Ca}$
for the potassium atoms and calcium ions, respectively, removes these resonances as shown in Fig.~\ref{rateXA}b).
The rate coefficient is no larger than $4 \times 10^{-18}$ cm$^3$/s, eight orders of magnitude smaller than
that found in Sec.~\ref{ssec:ioncool} and observed in our experiment.

\begin{figure}
 \includegraphics[width=\columnwidth,trim=0 0 0 0,clip]{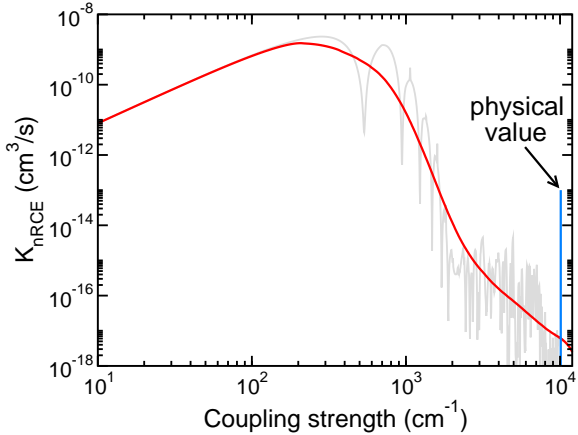}
 \caption{ 
 Non-radiative charge-exchange rate coefficients for
 K$(^{2}{\rm S}_{1/2})+$Ca$^{+}(^{2}{\rm S}_{1/2})$ to K$^{+}(^{1}{\rm S}_0)+$Ca$(^{1}{\rm S}_0)$ as 
 functions of diabatic coupling strength $U_{12}(R)$ at $R=9a_0$, the separation where the two diabatic potentials 
 cross  (See Fig.~\ref{diabatic}a) and d).) The gray curve corresponds to $K_{\rm nRCE}(E)$ at $E/k=0.2$ K, while
 the red curve corresponds to the rate coefficient based on  Landau-Zener
 theory.
 The nominal or physical coupling strength is $U_{12}(9a_0)/hc\approx 9500$ cm$^{-1}$
and is indicated by a blue vertical line.}
\label{LZrate}
\end{figure}

We finish the discussion of Pathway I by artificially changing the coupling function $U_{12}(R)$ in order
to better understand why our prediction for its charge-exchange rate coefficient is so small. We uniformly scale
$U_{12}(R)$ and have repeated the calculation of  $K_{\rm nRCE}(E)$  for various scale factors. Figure \ref{LZrate}
shows the rate coefficient at a single collision energy as  a function of $U_{12}(R)$
at $R=9a_0$, where the diabatic potentials cross, {\it i.e.} $U_{11}(R)=U_{22}(R)$.
The figure also shows a prediction based on the Landau-Zener curve-crossing model
\cite{Zhu1995,Coveney1985} using the coupling strength and value and slopes of the diabatized potentials at the crossing point.
Our numerical rate coefficient has a characteristic peaked behavior in good agreement with the
Landau-Zener theory. It is largest for $U_{12}(9a_0)/hc\approx200$ cm$^{-1}$
and rapidly decreases for both smaller and larger $U_{12}(9a_0)$.
The rate coefficient based on the Landau-Zener model does not capture the  oscillatory behavior of  $K_{\rm nRCE}(E)$ with coupling strength as
quantum interferences are not captured.
Clearly, for the physical coupling strength the rate coefficient is almost negligibly small.

\subsection{Radiative charge exchange by Pathway II}

The description of  charge-exchange   by spontaneous emission from K$(^{2}{\rm S}_{1/2})+$Ca$^{+}(^{2}{\rm S}_{1/2})$
collisions in pathway II involves the same Born-Oppenheimer potentials as in pathway I. In addition
to the potentials we require the electronic transition dipole moment between the two energetically-lowest
$^{1}\Sigma^{+}$ adiabatic states, $d_{{\rm XA},e}(R)$ as a function of $R$. 
Figure~\ref{fig:other}b) shows dipole moment $d_{{\rm XA},e}(R)$  determined by our MRCI calculations.
The transition dipole moment is largest near $R=10a_0$ close to the inner turning point of the excited A$^{1}\Sigma^{+}$
potential for our cold collision energies. It approaches zero for large separations.

We use the optical potential (OP) approach combined with the distorted-wave Born approximation for the phase shift~\cite{Zygelman1988,Sayfutyarova2013} to calculate the
total radiative charge-exchange rate coefficient $K_{\rm RCE}(E)=v\sigma_{\rm RCE}(E)$. The cross section $\sigma_{\rm RCE}(E)$ is then
\begin{equation}
\sigma_{\rm RCE}(E)=\frac{\pi}{k^2}g\sum_{\ell=0}^\infty (2\ell+1)
    \left\{1-e^{-4\eta_\ell(E)}\right\} \;,
\end{equation}
where wavenumber $k=\sqrt{2\mu E/\hbar^2}$ and 
the dimensionless quantity.
\begin{eqnarray}
\eta_\ell(E) &=& \frac{2\pi}{3}\frac{\mu}{k}\alpha^3\int^\infty_0 \mathrm{d}R
\,   |F_{\rm A}^{E\ell}(R)|^2
\label{eq:eta} \\
&& \quad\quad\quad\quad \times |d_{{\rm XA},e}(R)|^2 [V_{\rm A}(R)-V_{\rm X}(R)]^3  \;, 
\nonumber
\end{eqnarray}
where $\alpha\approx 1/137$ is the fine-structure constant and  $F_{\rm A}^{E\ell}(R)$ is
a solution of the single-channel radial Sch\"{o}dinger equation for the A$^1\Sigma^+$ potential
with collision energy $E$ and partial wave $\ell$. In fact, $F_{\rm A}^{E\ell}(R)\to \sqrt{2/\pi} \sin(kR+\delta_{\rm A,\ell}(k))$ for $R\to\infty$,
where $\delta_{\rm A,\ell}(k)$ is the phase shift of the A$^1\Sigma^+$ potential.  In Eq.~\ref{eq:eta}  all quantities are expressed in atomic units.
Finally, ${g=1/4}$ is a statistical probability that accounts for the fact that  the entrance threshold has  singlet and  triplet total electron spin channels
and that only the singlet $^1\Sigma^+$ channel leads to radiative transitions.  

\begin{figure}
 \includegraphics[width=\columnwidth,trim=0 0 0 0,clip]{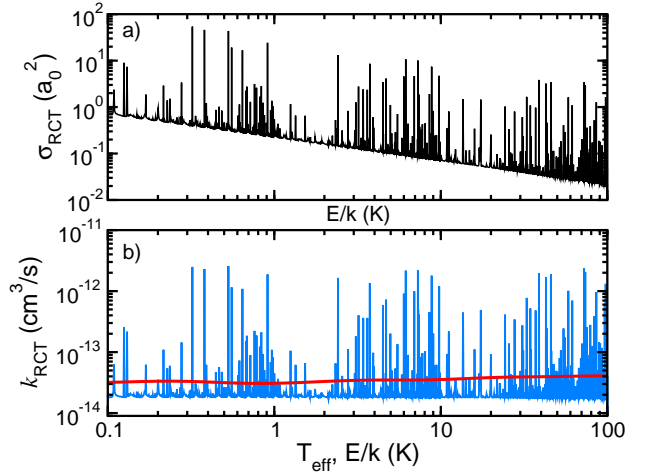}
 \caption{ Panel a: Radiative charge exchange cross-section from
 K$(^{2}{\rm S}_{1/2})+$Ca$^{+}(^{2}{\rm S}_{1/2})$ to K$^{+}(^{1}{\rm S}_0)+$Ca$(^{1}{\rm S}_0)$ 
 as a function of the entrance-channel collision energy. Panel b: Thermalized (ref line) and non-thermalized rate coefficients as a function of temperature and collision energy.}
\label{rateRad}
\end{figure}

Figure~\ref{rateRad}a) shows our radiative charge exchange cross section $\sigma_{\rm RCE}(E)$ for pathway II as a function of collision energy. 
The sharp resonances are again due to shape resonances of the induction potential  from the forty partial
waves that contribute significantly at $E/k=0.2$ K. Thermally averaging assuming temperatures $T_{\rm K}$ and $T_{\rm Ca}$
for the potassium atoms and calcium ions, respectively, removes these resonances as shown in Fig.~\ref{rateRad}b).
The rate coefficient is nearly independent of temperature and no larger than $4 \times 10^{-14}$ cm$^3$/s. This value is much larger than that for the non-radiative 
process in pathway I, but still four orders of magnitude smaller than
 observed in our experiment.
 
Finally, we estimate that the radiative charge-exchange rate coefficient with the reactants K$(^{2}{\rm P})$ and Ca$^{+} (^{2}{\rm S}$) is small, of the order of 10$^{-14}$ cm$^3$/s, based on a similar analysis as in Sec. III.C. Our experiment is not able to detect charge exchange rate coefficients of this order of magnitude. 
 
 \subsection{Non-radiative charge exchange with Pathway III} 
 
A potassium atom can  be resonantly excited to the 4p $^2$P$_{3/2}$ state by the absorption of a  
photon from the MOT lasers and then react with the calcium ion. That is,
the reactants are  ${\rm K}(^2{\rm P})$ and ${\rm Ca}^+(^2{\rm S})$. Unlike for Pathway I,
 more than one product state exists. Several of them have an electronic energy that is a few times 
$hc\times10^3$ cm$^{-1}$ lower in energy than that of the reactants and can lead to charge exchange (See figure in Methods.)
Based on the results for Pathway I and, in particular, those in Fig.~\ref{LZrate},  we limit ourselves
to the product state ${\rm K}^+(^1{\rm S}_{0})+{\rm Ca}(^1{\rm P}_1)$, which releases the least amount of relative 
kinetic energy. Even then, a model of charge exchange should include  $\Sigma$ as well as $\Pi$ molecular states/channels
and, as the atomic ${\rm K}(^2{\rm P})$ configuration is split into P$_{1/2}$ and P$_{3/2}$ states,
spin-orbit interactions. As our goal is to only obtain an order of magnitude estimate for the rate coefficient
it is sufficient to setup a model of charge exchange solely based on the singlet $^1\Sigma^+$ potentials.

 \begin{figure}
 \includegraphics[width=0.9\columnwidth,trim=0 5 0 20,clip]{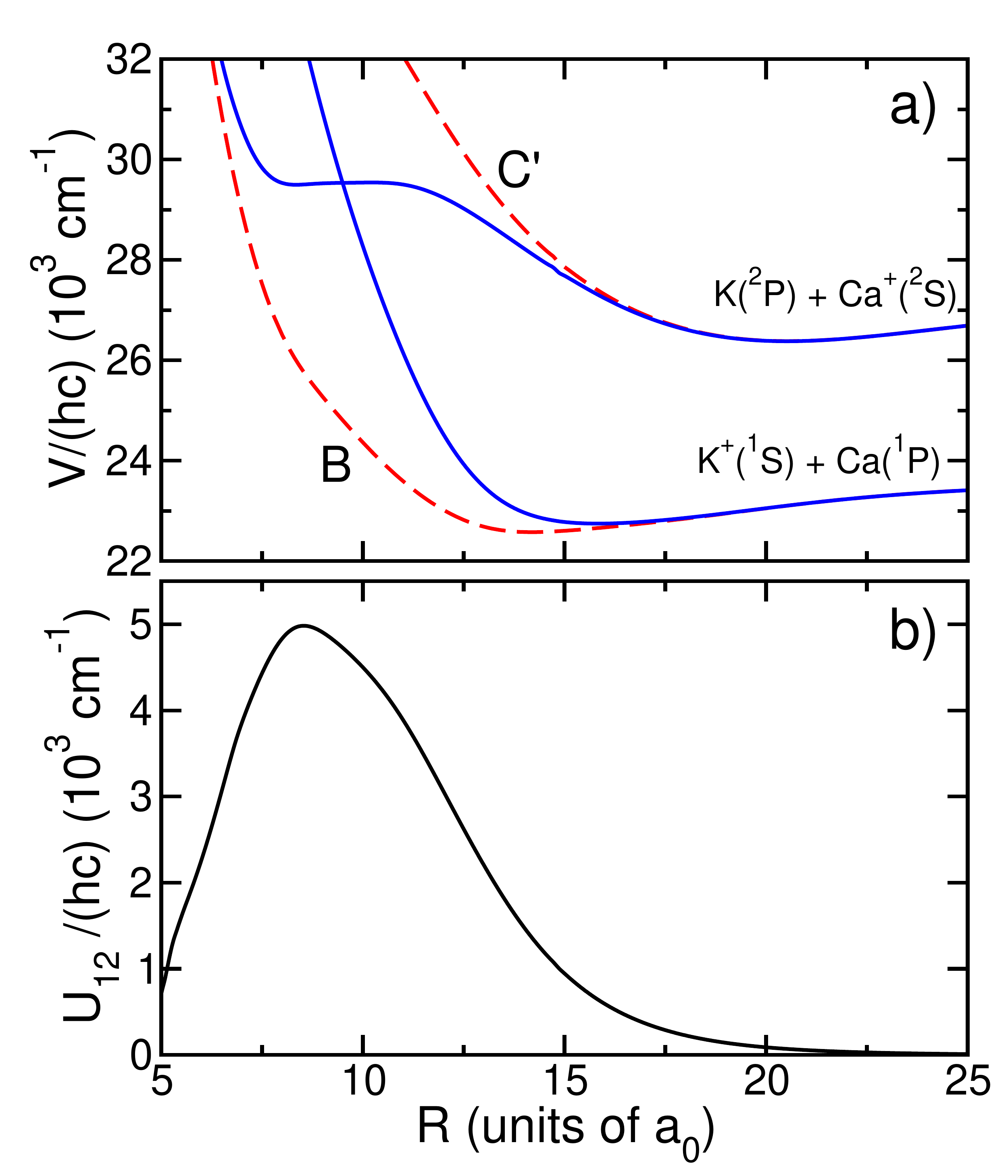}
 \caption{Potentials and couplings relevant for the  non-radiative
 ${\rm K}(^2{\rm P})+{\rm Ca}^+(^2{\rm S})  \to {\rm K}^+(^1{\rm S})+
{\rm Ca}(^1{\rm P})$ charge-exchange process. 
 (a) The excited B$^1\Sigma^+$  Born-Oppenheimer
 potential and partially-diabatized C$^\prime\,^1\Sigma^+$ potential, as defined in the text, from our MRCI calculations (dashed red lines)
 as functions of separation $R$.
 The derived diabatic potentials are shown as solid blue curves.
 (b) Coupling function $U_{12}(R)$ between the  
 diabatic  B and C$^\prime$  potentials as a function of $R$. 
 }
\label{diabatic1}
\end{figure}

Figure \ref{fig:other}a) shows the three relevant highly-excited  $^1\Sigma^+$  Born-Oppenheimer 
potentials, each dissociating to one of the ${\rm K}^+(^1{\rm S}_{0})+{\rm Ca}(^1{\rm P}_1)$, ${\rm K}(^2{\rm P})+{\rm Ca}^+(^2{\rm S})$,
and ${\rm K}(^2{\rm S})+{\rm Ca}^+(^2{\rm D})$ limits.  Avoided crossings between the potentials occur at multiple separations. 
We were, however, unable to compute the non-adiabatic coupling function $Q_{s,s'}(R)$ 
among the corresponding B$^1\Sigma^+$, C$^1\Sigma^+$, and D$^1\Sigma^+$ states within the MRCI method.
Instead, we  resort to further approximations based on the observation that
the splitting between the C$^1\Sigma^+$ and D$^1\Sigma^+$ potentials at the avoided crossing near $R=R_{\rm CD}\equiv 15a_0$ 
is  small compared to that of others at smaller separations. 
In fact, the splitting is much smaller than the energy difference between the C$^1\Sigma^+$ and D$^1\Sigma^+$ potentials at  $R=R_{\rm CD}$  and
that of the ${\rm K}(^2{\rm P})+{\rm Ca}^+(^2{\rm S})$ reactants and, classically, the crossing can not be accessed. This then implies that 
to good approximation we can construct a smooth diabatic potential $U_{ii}(R)$ that closely
follows the D$^1\Sigma^+$ potential for $R<R_{\rm CD}$ and the C$^1\Sigma^+$ potential for $R>R_{\rm CD}$.
We will denote  this potential by C$^\prime\,^1\Sigma^+$. The other diabatic potential
will not be relevant for charge exchange as it dissociates to the ${\rm K}(^2{\rm S})+{\rm Ca}^+(^2{\rm D})$ limit.
We omit the coupling between these two diabatic potentials.
 
Figure~\ref{diabatic1}a) shows the B$^1\Sigma^+$ and C$^\prime\,^1\Sigma^+$ potentials.
The potentials have a broad barely-visible avoided crossing between $7a_0$ and $13a_0$ similar in shape to that
between the X$^1\Sigma^+$ and A$^1\Sigma^+$ potentials for Pathway I. In other words, there exist a
weak non-adiabatic coupling between the {\it adiabatic} B$^1\Sigma^+$ and C$^\prime\,^1\Sigma^+$ states
and, in fact, we will assume that the mixing angle $\theta(R)$ between states B and C$^\prime$ is the same as that between the 
X and A states shown in Fig.~\ref{diabatic}c). The derived diabatic potentials 
$U_{11}(R)$, $U_{22}(R)$, and $U_{12}(R)$ are shown in Figs.~\ref{diabatic1}a) and \ref{diabatic1}b). 

An further analysis of the potentials in Fig.~\ref{diabatic1} shows that the 
inner turning point for the reactant collision with relative kinetic energies near $E/k=1$ K
is around $18a_0$ for both the adiabatic C$^\prime\,^1\Sigma^+$ potential
and the diabatic potential dissociating to the same limit.
Moreover, the energy of the diabatic potentials at the separation where they cross
is higher than that of the ${\rm K}(^2{\rm P})+{\rm Ca}^+(^2{\rm S})$ reactants.
The crossing is not accessible when $U_{12}(R)$ is either small or large. 
For the physically-appropriate shape of $U_{12}(R)$, the numerically solution of 
the two coupled channels gives a thermalized rate coefficient of less than $10^{-18}$ cm$^3$/s at $T_{\rm eff}=0.2$ K, smaller
than that for pathway I.

\section{Discussions}

We have presented our experimental and theoretical investigations of charge exchange interaction between cold $^{40}$Ca$^+$ ions and $^{39}$K atoms. Our theoretical analysis of all possible collision pathways confirms that the reaction occurs via the ${\rm K}(^2{\rm S}_{1/2})+{\rm Ca}^+(^2{\rm P}_{1/2})\rightarrow {\rm K}^+({\rm^1S}_0) + Ca(3d4p^3{\rm P}_1)+\hbar\omega$ channel.
The experimentally measured charge exchange rate coefficient is in good agreement with our theoretical calculations.

In the future, it is possible to induce non-adiabatic transition between the first and the second electronic potentials of KCa$^+$ molecules using an additional high intensity laser. One of the near future goal is to study the effect of this strong interaction on the charge exchange rate.

Over the past decade, hybrid traps proved to be a good platform to study the elastic as well as rich chemical interactions between cold atoms and ions. The next step ahead is to use this platform for attaining the ro-vibrational ground state of molecular ions. The laser cooled atomic ions can efficiently cool the translational degrees of freedom of the molecular ions \cite{Kimura2011,Rugango2015} and the internal states of the molecular ions can be sympathetically cooled using ultra cold atoms. Research towards this goal over the past few years shows promising results \cite{Rellergert2013,Hansen2014}. Our next goal involves introducing CaH$^+$ molecular ions to the hybrid apparatus and cooling their internal and external degrees of freedom using cold K atoms and Ca$^+$ ions. However, the undesired interactions between the atoms, ions and molecular ions \cite{Hudson2009} as well as unwanted dissociation \cite{Jyothi2016} of the molecular ions in the presence of lasers need to be controlled to achieve this goal. 
For instance, by manipulating the electronic state population of the $^{40}$Ca$^+$ ions we could control the K-Ca$^+$ interaction rate and increase the $^{40}$Ca$^+$ ions lifetime in the presence of K atoms from 2 s to 12 s. Internal state cooling of molecular ions can be observed in the time scale of milliseconds.
The already established resolved ro-vibrational spectroscopy measurement of CaH$^+$ \cite{Calvin2018} can be used for the detection of the internal cooling of CaH$^+$. Once prepared in the ground state, the CaH$^+$ molecular ions are well suited for the test of fundamental theories such as time variation of the proton to electron mass ratio \cite{Schiller2005}.

\begin{acknowledgments}
This work is supported by the MURI Army Research Office Grant
W911NF-14-1-0378-P00008 and ARO Grant W911NF-17-1-
0071.
\end{acknowledgments}

\providecommand{\noopsort}[1]{}\providecommand{\singleletter}[1]{#1}%
\providecommand*{\mcitethebibliography}{\thebibliography}
\csname @ifundefined\endcsname{endmcitethebibliography}
{\let\endmcitethebibliography\endthebibliography}{}

\end{document}